\documentclass[manuscript]{aastex}

\slugcomment{to appear in ApJ}

\shorttitle{Triggering Solar Flares}
\shortauthors{Kusano et al.}

\begin{document}

\title{Magnetic Field Structures Triggering Solar Flares and \\
Coronal Mass Ejections}

\author{K.\ Kusano\altaffilmark{1}, Y.\ Bamba, and T.\ T.\ Yamamoto}
\affil{Solar-Terrestrial Environment Laboratory, 
Nagoya University, Furo-cho, Chikusa-ku, Nagoya, 
Aichi, 464-8601, Japan}

\author{Y.\ Iida\altaffilmark{2}, and S.\ Toriumi\altaffilmark{2}}
\affil{Department of Earth and Planetary Science, University of Tokyo, Hongo, 
Bunkyo-ku, Tokyo, 113-0033, Japan}

\author{A.\ Asai}
\affil{Unit of Synergetic Studies for Space, Kyoto University, 17 Kitakazan Ohmine-cho, Yamashina-ku, Kyoto, 607-8471, Japan}

\email{kusano@nagoya-u.jp}

\altaffiltext{1}{Japan Agency for Marine-Earth Science and 
Technology (JAMSTEC), 
Yokohama, Kanagawa 236-0001, Japan}
\altaffiltext{2}{JSPS Research Fellow}

\begin{abstract}
Solar flares and coronal mass ejections (CMEs),
the most catastrophic eruptions in our solar system,
have been known to affect terrestrial environments 
and infrastructure. However, because their
triggering mechanism is still not sufficiently understood,
our capacity to predict 
the occurrence of solar eruptions and to forecast space weather
is substantially hindered.
Even though various models have been proposed to determine the onset of 
solar eruptions, the types of
magnetic structures capable of 
triggering these eruptions are still unclear.  In this study,
we solved this problem by systematically surveying the nonlinear 
dynamics caused by a wide variety of magnetic structures in terms of 
three-dimensional magnetohydrodynamic 
simulations. As a result, we determined that two 
different types of small magnetic structures  
favor the onset of solar eruptions.
These structures, 
which should appear near the magnetic polarity inversion 
line (PIL), include magnetic fluxes reversed
to the potential component or the nonpotential component 
of major field on the PIL.
In addition, we analyzed two large flares, the X-class flare on December 13, 
2006 and the M-class flare on February 13, 2011, using 
imaging data provided by the Hinode satellite, and we demonstrated that 
they conform to the simulation predictions.
These results suggest that forecasting of solar eruptions
is possible with sophisticated observation of a solar
magnetic field, although
the lead time must be limited by the time scale of changes in 
the small magnetic structures. 
\end{abstract}



\section{Introduction}

Solar flares and coronal mass ejections (CMEs), the
most catastrophic eruptions in our solar system, are believed to 
occur through a type of storage-and-release phenomena, 
in which the free energy 
stored as a nonpotential magnetic field in the solar corona 
is abruptly unleashed \citep{priest_forbes_2002, shibata_magara_2011}. 
It is a promising hypothesis that the onset of
solar eruptions is related to some typical magnetic structures.
In fact, many studies have emphasized a relationship between 
the occurrence of solar eruptions and 
various magnetic properties; {\it e.g.} 
strong magnetic shear \citep{low_1977, hagyard_etal_1984, 
mikic_linker_1994, kusano_etal_1995, zhang_2001},
reversed magnetic shear \citep{kusano_etal_2004, wang_etal_2004, 
chandra_etal_2010, romano_zuccarello_2011, vemareddy_etal_2012},
sigmoidal structure of the coronal magnetic field 
\citep{rust_kumar_1996, canfield_etal_1999, moore_etal_2001},
flux cancellation on the photosphere \citep{van_ballegooijen_martens_1989, 
zhang_etal_2001, green_etal_2011}, 
converging foot point motion \citep{inhester_etal_1992, forbes_priest_1995, 
amari_etal_2003a, amari_etal_2003b, cheng_etal_2010},
the sharp gradient of magnetic field \citep{schrijver_2007},
emerging magnetic fluxes \citep{heyvaerts_etal_1977, moore_roumeliotis_1992, 
feynman_martin_1995, chen_shibata_2000, fan_gibson_2004,
chifor_etal_2007, wallace_etal_2010, moore_etal_2011,
archontis_hood_2012},
multipolar topologies \citep{antiochos_etal_1999, 
sterling_moore_2004, williams_etal_2005},
flux rope \citep{forbes_priest_1995, torok_kliem_2005, chen_1996, 
fan_gibson_2007},
narrow magnetic lanes between major sunspots 
\citep{zirin_wang_1993, kubo_etal_2007},
topological complexity \citep{schmieder_etal_1994},
intermittency and multifractality \citep{abramenko_yurchyshyn_2010},
and double loop structure \citep{hanaoka_1997}.

However, although some observations suggested that these properties 
tend to correlate with the solar eruption productivity,
the triggering mechanism for solar eruptions is still debated 
\citep{forbes_2000}, and
we do not yet well understand what determines
the onset of solar eruptions
\citep{leka_barnes_2007, welsch_etal_2011}.
It substantially hinders our capacity to predict 
the occurrence of solar eruptions and to forecast space weather.
The objective of this paper is to clarify the types of 
magnetic field structures that are 
capable of generating solar eruptions to gain a  
better understanding of their triggering mechanism.

Because the magnetic structures of eruption-productive active regions 
(ARs) are highly complex, it is difficult to determine 
the key elements of the magnetic field present at 
the onset of eruptions from the imaging data.
In this study, we compensate for this difficulty by 
systematically surveying the nonlinear 
dynamics caused by a wide variety of magnetic structures, 
which consists of a large scale force-free field and a
small scale bipole field,  
in terms of 
three-dimensional magnetohydrodynamics (MHD) 
simulations.
As a result, we show that two typical 
magnetic structures favor the onset of solar eruptions, 
and we demonstrate that
the simulation results are in good agreement with the 
observations with the Hinode satellite 
\citep{kosugi_etal_2007}.

The paper is organized in the following manner. 
The model and results of simulations
are described in Sections 2 and 3, respectively. 
On the basis of the simulation study, 
we propose the two scenarios for the triggering process of solar eruptions.
In Section 4, we show that two large flares observed by Hinode, 
the X-class flare on December 13, 
2006 and the M-class flare on February 13, 2011,
conform to the predictions determined through simulation.
Finally, in Section 5, we discuss the topological features of the magnetic 
field in the triggering phase of solar eruptions 
as well as the predictability of these eruptions
on the basis of this paper's conclusion.

\section{Simulation Model}

The simulation box includes a rectangular domain
$(-3L, -0.75L, 0) \leq (x,y,z) \leq (3L, 0.75L, 3L),$
which corresponds to the solar atmosphere above the photosphere within an AR. 
$L$ is the typical length and 
the coordinate $z$ represents the altitude from the photosphere.
The $x$ axis ($y=z=0$) corresponds to the magnetic polarity inversion line 
(PIL), on which the vertical component of magnetic field $B_z$ changes sign.

To model the preflare state, 
the initial magnetic field is given by the linear force-free field, 
\[ \mathbf{B}_{init} (\alpha; y, z) 
= B_0 \left( \begin{array}{@{\,}c@{\,}}
             \alpha k^{-1} \cos ky \\
                 -K k^{-1} \cos ky \\
                           \sin ky 
             \end{array}
      \right) \exp (-Kz),
\]
where $k=2 \pi/L,$ $K=(k^2-\alpha^2)^{1/2},$ and 
$B_0$ is a constant. This field is characterized by the scalar parameter
$\alpha$ and shear angle $\theta_0 = \tan^{-1}(-B_x/B_y)=\tan^{-1} \alpha/K,$
which is defined as the azimuthal rotation of the magnetic field 
with respect to the potential magnetic field $\mathbf{B}_{init}(0;0,0) = (0,-B_0,0),$ as shown in Fig.\ 1.

After the simulation starts, we quickly inject a small bipole magnetic field
into the force-free field $ \mathbf{B}_{init} $ from the bottom boundary
in order to form the partially disturbed magnetic structure
illustrated in Fig.\ 1. 
This process is performed by prescribing the boundary condition 
of electric field, 
which is generated by the constant ascending motion of 
a synthetic magnetic torus located 
virtually below the simulation box. 

The ascending torus forms a sphere of 
radius $r_e$, within which there is 
only the toroidal magnetic field $\mathbf{B}_{e}$ of constant intensity. 
The major axis of the torus is on the $x-y$ 
plane and its center is initially at a point $(0, y_e, -r_e).$ 
The torus ascends with constant velocity $v_e$ 
for only the period $0 \leq t  \leq \tau_e ( \leq r_e/v_e).$ 
The electric field  $\mathbf{B}_e \times \mathbf{v}_e$ 
and velocity $\mathbf{v}_e$ are imposed within 
the cross-section of the ascending torus on the bottom plane until $t = \tau_e.$ The injected small scale field is 
characterized by the azimuthal orientation $\varphi_e$ of $\mathbf{B}_e$ 
at the top of torus, the offset of torus center $y_e,$ 
and the total amount of injected flux that is a function of $B_e,$ $r_e,$ and  
$v_e \tau_e$ (see Fig.\ 1). 

Although this model corresponds to the kinematic model of 
emerging flux \citep{fan_gibson_2004}, we adopt it
as a way to dynamically form a variety of magnetic structures 
rather than as the model of flux emerging.
In fact, because the speed of small magnetic field injection
is about two orders of magnitude faster than the realistic speed 
of emerging flux (the order of 1 km/s), 
the physically meaningful process of simulation 
is after the flux injection terminates. 
Therefore, we focus mainly on the dynamics caused by
magnetic structures formed as the result of flux injection 
rather than the dynamics of flux emerging.
Which process in flux emerging or horizontal motion 
on the solar surface is more efficient for triggering the solar eruptions
is beyond the scope of this paper, although we partially discuss 
this issue in Section 3.

The length, magnetic field, and time are described by non-dimensional 
units, $L,$ $B_0,$ and the 
Alfv\'{e}n transit time $(\tau_A=L/V_A),$ respectively, 
where $V_A$ is the Alfv\'{e}n speed $B_0/\sqrt{\mu_0 nm_i}$  for ion 
mass $m_i,$ plasma number density $n$ and vacuum permeability $\mu_0.$ 
In a typical AR ($B_0 \sim 0.05\mbox{ T,}$ $n \sim 10^{15} \mbox{ m}^{-3},$ 
$L \sim 20 \mbox{ Mm}),$ $\tau_A \sim 0.4 \mbox{ s}$ and $V_A \sim 50 
\mbox{ Mm/s}.$ 

We adopted the zero-beta model for the simulation, in which plasma pressure 
and variations in plasma density are assumed to be less effective. This model 
is applicable when the plasma beta (i.e., the ratio of thermal to magnetic 
pressure) is significantly less than unity. The coronal plasma 
within an AR in the preflare phase, which is the focus of this paper, 
is thus covered by this model. 
The governing equations are the same as those used in our previous study
\citep{kusano_etal_2004, kusano_2005}. The electrical 
resistivity $\eta$ and viscosity $\nu$ are initially constant 
($10^{-5}$ and $5\times10^{-3}$ in non-dimensional units, 
respectively). However, $\eta$ increases if the electric current density 
$J$ exceeds a critical $J_c,$ as adopted 
from Eq.\ 9 of \citet{kusano_etal_2004}, 
where the enhanced resistivity $\eta_1=5 \times 10^{-4}$ and critical electric 
density $J_c = 50.$ 

The spatial differential operator is approximated by the second-order 
accurate finite difference method using three-grid-point stencils, 
and the temporal integration was performed using the 
Runge-Kutta-Gill method with fourth-order accuracy. 
The grid number included in the 
simulation box is $1024 \times 256 \times 512$ for each dimension 
$x,$ $y,$ and $z,$ and they are packed near the 
PIL so that the finest grid sizes are 
$\Delta x = 5.9 \times 10^{-3},$ 
$\Delta y = 2.9 \times 10^{-3},$ and 
$\Delta z = 5.9 \times 10^{-3},$ respectively. 

\section{Simulation Results}

We chose the shear angle $\theta_0$ of large scale force-free field
and the azimuthal orientation of the small scale injected field 
$\varphi_e$ as the parameters, and surveyed the parameter space 
$(0,0) \leq (\theta_0, \varphi_e) \leq (85^{\circ},360^{\circ})$ with 106 runs. 
The cases $\varphi_e=0^{\circ}$, $0^{\circ}<\varphi_e<180^{\circ}$, $\varphi_e=180^{\circ}$, and 
$180^{\circ}<\varphi_e<360^{\circ}$ correspond to the small scale field 
of the right 
polarity, normal shear, opposite polarity (OP), and reversed shear (RS), 
respectively, compared with the preexisting large scale field. 

The simulation results are summarized in Fig.\ 2, 
which clearly shows that the kinetic energy produced by 
eruption strongly depends on $\theta_0$, hence a large-scale eruption 
is possible only in strongly sheared cases ($\theta_0>75^{\circ}$). 
This result agrees well with the observations
\citep{hagyard_etal_1984}, and is logical because
a more strongly sheared field stores a greater amount of free energy as the
nonpotential component of the magnetic field. 
However, the most notable feature of this figure is that 
the occurrence of solar eruption is sensitive to $\varphi_e$. 
In particular, the eruption-producing cases, 
represented by diamonds in Fig.\ 2, exist mainly for 
$\varphi_e=123^{\circ}$ to $270^{\circ}$. 
This result indicates that strong shear alone is not a sufficient
condition for the onset of eruptions and that the occurrence of 
eruptions is governed by small magnetic structures.
The simulation results predict that the OP and RS-type
magnetic structures are capable of triggering solar eruptions. 

We detected a clear difference in the morphologies of magnetic fields
between cases triggered
by OP and RS-type configurations, 
represented in Fig.\ 2 by pink and blue diamonds, respectively. 
The typical dynamics of eruption caused by 
the OP-type field are explained in the following steps, as shown in Fig.\ 3: 
(1) Two sheared magnetic field lines
rooted near the small bipole of OP (blue tubes in Fig.\ 3a)
are reconnected via the bipole of OP
to form twisted flux ropes (green tubes in Figs.\ 3b-3d),
(2) after twisted flux ropes grow (Fig.\ 3e), they 
suddenly erupt upward (Fig.\ 3f), 
(3) in the strongly sheared case ($\theta_0 > 70^{\circ}$),
the eruption of twisted flux ropes vertically stretches 
the overlying field lines, and new vertical
current sheets are generated below them (red surface in Fig.\ 3f), and
(4) magnetic reconnection begins on this
vertical current sheet and forms a cusp-shaped post-flare arcade along with 
more twisted ropes, which accelerate the eruption (Fig.\ 3g). 
In particular, for the most
strongly sheared case ($\theta_0 > 75^{\circ}$), steps (3) and 
(4) reinforce each other and the reconnection region
propagates along the PIL (Fig.\ 3h) until reaching the boundary 
of the simulation box, which corresponds
to the outer border of the AR. 

As a result, in a more sheared arcade, greater kinetic energy is
produced by the ascent of longer flux ropes. This process is 
essentially the same as the model of tether cutting with emerging flux
reported by \cite{moore_roumeliotis_1992}. 
The rapid ascent of the twisted rope can be attributed to the loss of
equilibrium and ideal MHD stability 
\citep{forbes_priest_1995, kliem_torok_2006, demoulin_aulanier_2010}.


On the other hand, the RS-type flux 
triggers eruptions in a different manner. As seen in Fig.\ 4, the
injected bipole field of reversed shear 
contacts the preexisting sheared fields
(blue tubes in Figs.\ 4a-4c) and
forms a current sheet (red surface denoted by {\sf c} in Fig.\ 4b) on 
the border between them.
Magnetic reconnection slowly proceeds on this
current sheet, and the sheared field is removed from the 
center to the sides of RS field region
({\sf d} and {\sf d'} in Fig.\ 4b). 
Because of the reduction of the sheared field in the center,
the magnetic arcade above the injected small bipole
partially collapses into the center, and a vertical 
current sheet ({\sf v} in Fig.\ 4d) is generated.
As a result, sheared magnetic fields
(sky-blue tubes in Fig.\ 4d) are reconnected 
to form twisted flux ropes rooted on {\sf F} and {\sf F'} in Fig.\ 4e. 
These twisted flux ropes grow (Fig.\ 4f) 
and erupt upward (Fig.\ 4g).
Finally, the processes continue in the same manner as 
those in steps (3) and (4) for the OP-type case
(Fig.\ 4h), and the eruption develops.

This trigger scenario is essentially the same as the reversed-shear 
flare model proposed by \citet{kusano_etal_2004}. However, 
it should be noted that
they studied the process that horizontal flow along the PIL, 
rather than flux emerging, reverses magnetic shear.
Therefore, we can conclude that
emerging flux is not necessarily needed for the onset of
eruption, if horizontal motion drives the formation of the 
magnetic structures favorable to the onset of 
eruptions.

It should be noted that the OP-type configuration may cause an eruption of 
flux rope before the onset of flare
reconnection, whereas the RS-type configuration triggers flare 
reconnection before the eruption
of flux rope. We refer to these processes as
``eruption-induced reconnection'' 
and ``reconnection-induced eruption,''
respectively. Figure 2 indicates that
while the eruption-induced reconnection is possible for any value of 
sheared angle $\theta_0$, reconnection-induced eruption 
can occur only for relatively large $\theta_0$. 
We can easily explain this result because 
some amount of magnetic shear has to pre-exist for reconnection
of the RS field to cause 
partial collapse of the magnetic arcade.

In both processes triggered by the OP and RS-type configurations, 
the morphology of the magnetic field in the latter phase 
(after the onset of flare reconnection) is
common and consistent with the standard CSHKP model of 
a two-ribbon flare \citep{carmichael_1964, sturrock_1966, hirayama_1974, 
kopp_pneuman_1976}. 
However, the initiation procedure differs between the two configurations.
It indicates that although both the eruption of flux rope,
which may create CMEs, and the reconnection of flares can
induce each other, the causality between them may be 
governed by small magnetic structures that 
trigger the eruptions.

\section{Comparison with Observations}

On the basis of our simulations, we determined that the 
magnetic structures favorable to the onset of
large solar eruptions consist of the OP or RS-type small 
magnetic fields 
and the strongly sheared field, as illustrated in Fig.\ 5.
For either type of trigger processes, 
major flares should be preceded by minor
reconnection between the preexisting sheared field and 
the flux element with a different orientation. 
These results are consistent with the observations of internal reconnection
\citep{moore_etal_2001}.
Therefore, we can anticipate preflare brightening related to the precedent 
reconnection in both OP or RS-type regions as a
precursor to major eruptions. In fact, we have determined that two major 
flares observed by the Solar Optical Telescope (SOT) 
\citep{tsuneta_etal_2008} aboard Hinode 
conform to this prediction.

\subsection{Case of 2006 December 13 Flare}
The first event is the X3.4-class flare observed in AR NOAA 10930 
at 02:14 UT on December 13, 2006 \citep{kubo_etal_2007}. 
According to the vector 
magnetogram observed by the SOT/Stokes Spectro Polarimeter (SP) installed 
on Hinode, the azimuthal 
magnetic field in this region was mainly twisted in a clockwise 
direction by more than $75^{\circ}$.
We analyzed the spatiotemporal 
correlation between the line-of-sight component of the
magnetic field and the Ca II H line emission,
which were observed by Narrowband Filter Imager (NFI)
and Broadband Filter Imager (BFI) of SOT, respectively. 
As evident in Fig.\ 6a, 
a small isolated positive pole (denoted by {\bf I}) 
slowly grew since 23:00 UT on December 12.
According to \citet{kubo_etal_2007}, 
this small magnetic flux emerged around 0 UT on December 12 at
the west side of the positive umbra, and it 
was elongated by the rotating flows around the southern positive umbra.
As a consequence, an OP-type 
configuration was formed in the region denoted by a yellow circle
(Fig.\ 6b), 
in which the orientation of a small bipole 
(positive in the north and negative in the south)
was opposite to the orientation of positive and negative major spots,
as illustrated in the subset at the lower right corner of Fig.\ 6b. 
A brightening of the Ca II H line began on the PIL of 
this OP-type bi-pole 30 min prior 
to the flare onset (Fig.\ 6b) and extended to 
the flare ribbons (Figs.\ 6c-6i).

In the very early stage of the flare (Figs.\ 6c and 6e), 
the bright kernels consisted of
sheared ribbons ({\sf F} on the positive pole; {\sf F'} on the negative pole) and
small segments ({\sf P$_1$}, {\sf P$_2$}, and {\sf P$_3$}).
Because the chirality of sheared ribbons is consistent with the observation 
of negative magnetic helicity in this region \citep{park_etal_2010},
the sheared ribbons can be interpreted as the feet
of twisted flux rope formed by reconnection in the
OP-type trigger region as seen in Figs.\ 3b and 3c. 
Here it should be noted that the feet {\sf F} and {\sf F'} in Fig.\ 3b corresponds to a mirror image of the 
observed {\sf F} and {\sf F'} in Fig.\ 6c because the sign of preloaded 
magnetic helicity differs. The sign is negative in this AR and positive in the simulation.

The transient bright segments ({\sf P$_1$}, {\sf P$_2$}, and {\sf P$_3$})
are consistent with the simulation results (Fig.\ 3),
assuming that segments appeared on the 
intensive electric current sheets formed on the top, 
bottom and lateral side of the twisted rope indicated as
{\sf S$_1$}, {\sf S$_2$}, and {\sf S$_3$}, respectively, in Figs.\ 3b-3d.
The current sheet {\sf S$_1$} is formed between the twisted rope and overlying 
field represented by blue tubes in Fig.\ 3d. 
Therefore, if a chromospheric emission {\sf P$_1$}
corresponds with the current sheet {\sf S$_1$}, it implies that the 
lowest part of the twisted rope, which forms a bald patch 
\citep{titov_etal_1993}, existed in the chromosphere. 
The fact that the bright segment {\sf P$_1$} quickly disappeared 
at 02:20 UT (Fig.\ 6e) can be attributed to the ascension of the
flux rope beyond the chromosphere. 
In the soft X-ray images observed by X-Ray Telescope (XRT)
\citep{golub_etal_2007} installed on Hinode,
a faint ejection is visible from 02:18:18 to
02:22:18 UT, as indicated by arrows in Figs.\ 6f and 6h.
This faint ejection exhibited a bar-like structure along the magnetic
neutral line and the ejection speed changed from 40 to 100 km/s.
It is likely that the ejection observed by XRT corresponds to 
the front of the flux rope launched from the chromosphere to the corona. 

While the {\sf P$_2$} segment was weakened (Figs.\ 6c, 6e, and 6g), 
the {\sf P$_3$} segment grew and was combined with ribbon {\sf F} (Fig.\ 6g).
In addition, these short-term evolutions of the bright segments
agree well with the simulation result in which the 
current sheet {\sf S$_2$} is weakened; however,  
{\sf S$_3$} grows
to the major current sheet below the erupting flux rope (Figs.\ 3d-3f). 
Finally, the ribbon on the negative pole ({\sf F'} in Fig.\ 6e) 
was extended eastward
to form ribbon {\sf R$_2$} and the ribbon on the positive pole {\sf F} 
thickened, as denoted by {\sf R$_1$} in Figs.\ 6i and 6j. 
The structure and dynamics of
these main two-ribbons are consistent with the 
evolution of the feet of the post-flare arcade observed 
in the simulation results ({\sf R$_1$} on the positive pole
and the counterpart of 
{\sf R$_2$'} on the negative pole in Figs.\ 3g and 3h).
From this excellent agreement with the observation and simulation,
we can conclude that the onset of this event can be explained 
by the OP-type trigger model.


\subsection{Case of 2011 February 13 Flare}
The second event, which can be explained by the RS-type trigger model, 
is the M6.6-class flare observed in AR NOAA 
11158 at 17:28 UT on 
February 13, 2011 (Fig.\ 7). The vector magnetogram (SOT/SP) 
indicates that the magnetic field in this 
region was also extensively twisted 
(more than $75^{\circ}$) in a counterclockwise 
direction. This counterclockwise twist, which corresponds to the 
preloaded positive magnetic helicity,
is consistent with the
sheared structure of the flare ribbons (Fig.\ 7d). 

Here we emphasize that an inverse S-shaped structure of the PIL, 
in which positive and negative polarity 
regions edge into each other, was formed 
in the region denoted by yellow circles in Figs.\ 7a-7c
between 13:15 UT and 15:00 UT. 
On the center of this region, the PIL is running north-south, 
whereas the overall PIL generally is running east-west. 
Therefore, it is highly likely that magnetic field crossing
the inverse S-shaped PIL is reversely directed 
compared with the large scale sheared magnetic field, and that
the RS-type configuration 
is formed as illustrated in the subset at the lower right corner 
of Fig.\ 7b.

In fact, the 
brightening of the Ca II H line ({\sf P$_1$} in Fig.\ 7b)
was observed since 15:00 UT in this area.
This preflare brightening can be attributed to 
heating of the current sheet
between the pre-existing sheared field and magnetic field of RS,
which corresponds to {\sf c} in Fig.\ 4b.
Immediately after the onset of the flare, 
sheared ribbons {\sf F} and {\sf F'} appeared as a center on the 
inverse S-shaped PIL (Fig.\ 7d).
The ribbon structure is highly consistent with the simulation result
such that the feet of the twisted ropes ({\sf F} and {\sf F'} in Fig.\ 4) and 
post-flare arcades ({\sf R} and {\sf R'}) combined to create elongated
sheared ribbons on both sides of the RS field region
(Fig.\ 4h).

Furthermore, it is worthwhile to note that the preflare brightening
({\sf P$_1$} in Fig.\ 7) was weakened before the onset of the flare 
(see Fig.\ 7c). 
In the simulation, the electric current intensity on the current sheet 
{\sf c} was weakened just prior to 
eruption of twisted ropes (see Fig.\ 4d).
This weakening resulted from the reconnection on the current sheet {\sf v}, 
because the reconnection ejected downward the shear-free field, which 
prevented contact between
the pre-existing sheared field and the field of RS.
The agreement of the observation and simulation results
indicates that the weakening of preflare brightening on the RS-like
region could be regarded as a 
precursor of flare onset.
 
From the magnetic structure and evolution of 
preflare and flare ribbons, we can conclude that this event 
is consistent with the RS-type trigger scenario. 
The formation process of the inverse S-shaped PIL in this region
is an interesting subject and we are analyzing the cause of 
the strucutre.
The detail results of analysis will be reported elsewhere. 

\section{Discussion and Conclusion}

The results shown in the previous section 
indicate that events indeed occurred, 
which are described by the following two 
scenarios:
Scenario 1 involves an eruption-induced reconnection process 
triggered by an OP-type magnetic field, and Scenario 2 is
a reconnection-induced eruption process 
triggered by an RS-type magnetic field.
Here, we examine the reason for existence of both scenarios.
It is explicable in terms of the topology of the magnetic field.
In both scenarios, flare reconnection on
a vertical current sheet is caused by the diverging flows that
remove magnetic flux and plasma from the reconnection site.
This mechanism corresponds to the ``pull'' mode in reconnection experiments
performed in laboratories \citep{yamada_1999}. 
As shown in Fig.\ 8, two different types of 
divergent flows are able to form thin current sheet {\sf v} 
between two sheared
magnetic loops {\sf B$_1$} and {\sf B$_2$}.
One divergent flow is a vertical (upward) flow above the PIL 
({\sf u} in Fig.\ 8), 
and the other is a horizontal flow along the PIL ({\sf h} and {\sf h'}). 
They correspond to the preflare flows that trigger
flare reconnection in Scenarios 1 and 2, 
respectively.
Because a divergent flow can flow either in the vertical or horizontal
direction, two different scenarios exist.

On the other hand, the converging flow ({\sf c} and {\sf c'})
can also form a vertical current sheet
and may drive flare reconnection, which  
corresponds to the ``push'' mode
in reconnection experiments \citep{yamada_1999}.
In fact, the converging flow on the solar surface
is believed to play the role of a trigger in
the flux cancellation (FC) model \citep{inhester_etal_1992,
priest_etal_1994, amari_etal_2003a}.
Therefore, from the perspective of magnetic topology, 
we can conclude that these three 
scenarios (Scenarios 1 and 2 proposed in this paper 
as well as the reconnection scenario
caused by the converging flow) 
can all apply to the triggering of solar flares.

The classification of the PIL morphology in flaring sites is
a promising technique used to examine the applicability of each scenario
because each scenario requires different structures of PIL 
as illustrated in Fig.\ 5.
Scenario 1 requires an OP-type disturbance to the PIL
(Fig.\ 5a).
On the other hand, 
the inverse S-shaped PIL (RS+ in Fig.\ 5b)
and the S-shaped PIL (RS$-$ in Fig.\ 5b) 
are the notable signature of the RS structures 
when the large-scale sheared field is positive
(counterclockwise rotated) and negative
(clockwise rotated), respectively, 
although an RS structure without
a disturbed PIL is possible \citep{kusano_etal_2004}.
In fact, such types of the PIL morphology
have been observed in several flare-productive active regions;
e.g. 
ARs NOAA 6659 (Fig.\ 1a of \citet{zirin_wang_1993})
and NOAA 9026 (Fig.\ 4q of \citet{kurokawa_etal_2002})
for an OP-type morphology,
AR NOAA 9026 (Fig.\ 4k of \citet{kurokawa_etal_2002})
for an RS-type morphology.
 
On the other hand, the FC model caused by the converging flow
does not require any disturbances
to the PIL if positive and negative magnetic fluxes cancel each other.
However, because the emergence of a magnetic element
in the flaring site is often associated with the 
cancellation of magnetic 
flux \citep{wang_shi_1993}, 
we should carefully recognize the structure of flare triggering.
Our analyses suggest that a high cadence 
(less than a couple of minutes) of preflare brightening observation 
in the chromospheric lines such as Ca II H 
is a powerful tool to recognize preflare processes.
The statistical study of various solar eruptions will explain
the applicability of each scenario to the onset of solar
flare and CMEs.

Our study suggests that 
solar flares and CMEs are triggered by magnetic 
disturbances in regions where magnetic shear has stored 
free energy rather than originating in 
phenomena that inevitably occur as a consequence 
of free energy storage; i.e., both magnetic shear 
(or magnetic helicity) and ``proper'' disturbance of the magnetic field
are necessary for the occurrence of solar flares and CMEs.
This conclusion 
implies that the lead time for flare and CME forecasting 
is limited by the time scale of changes 
in small magnetic elements that work as a trigger. 
Our analysis of the two events suggests 
a lead time shorter than several hours. 
Longer forecasting can be 
difficult and may be possible only in a probabilistic manner. 

While the simulations indicate that the orientation of 
a small magnetic element that perturbs the sheared magnetic
structure is crucial for triggering flares and CMEs, it is 
highly likely that the size and location (displacement from PIL) of
magnetic perturbation are also important.
The results of more advanced simulations to determine
the significance of these parameters will be published 
elsewhere.

\acknowledgments

We would like to thank Brian Welsch, Shinsuke Imada, Louise K.\ Harra, 
Bernhard Kliem, and Daikou Shiota for useful 
discussions. We are grateful to the referee for his/her appropriate and constructive comments to improve the paper. This work was supported by a Grants-in-Aid for Scientific Research (B) 
``Understanding and Prediction of Triggering Solar Flares''
(23340045, Head Investigator: K.\ Kusano) from the Ministry of Education, 
Science, Sports, Technology, and Culture of Japan. 
Simulations were conducted on the Earth Simulator in Japan Agency 
for Marine-Earth Science 
and Technology (JAMSTEC). Hinode is a Japanese mission developed 
and launched by 
ISAS/JAXA in collaboration with NAOJ as a domestic partner, 
and NASA and STFC (UK) as 
international partners. Scientific operation of the Hinode mission 
is conducted by the Hinode 
science team organized at ISAS/JAXA, which mainly consists of 
scientists from institutes in 
partner countries. Support for the post-launch operation is 
provided by JAXA and NAOJ 
(Japan), STFC (UK), NASA (USA), ESA, and NSC (Norway).

\clearpage
\begin{figure}[h] 
\begin{center}
\includegraphics[scale=0.9]{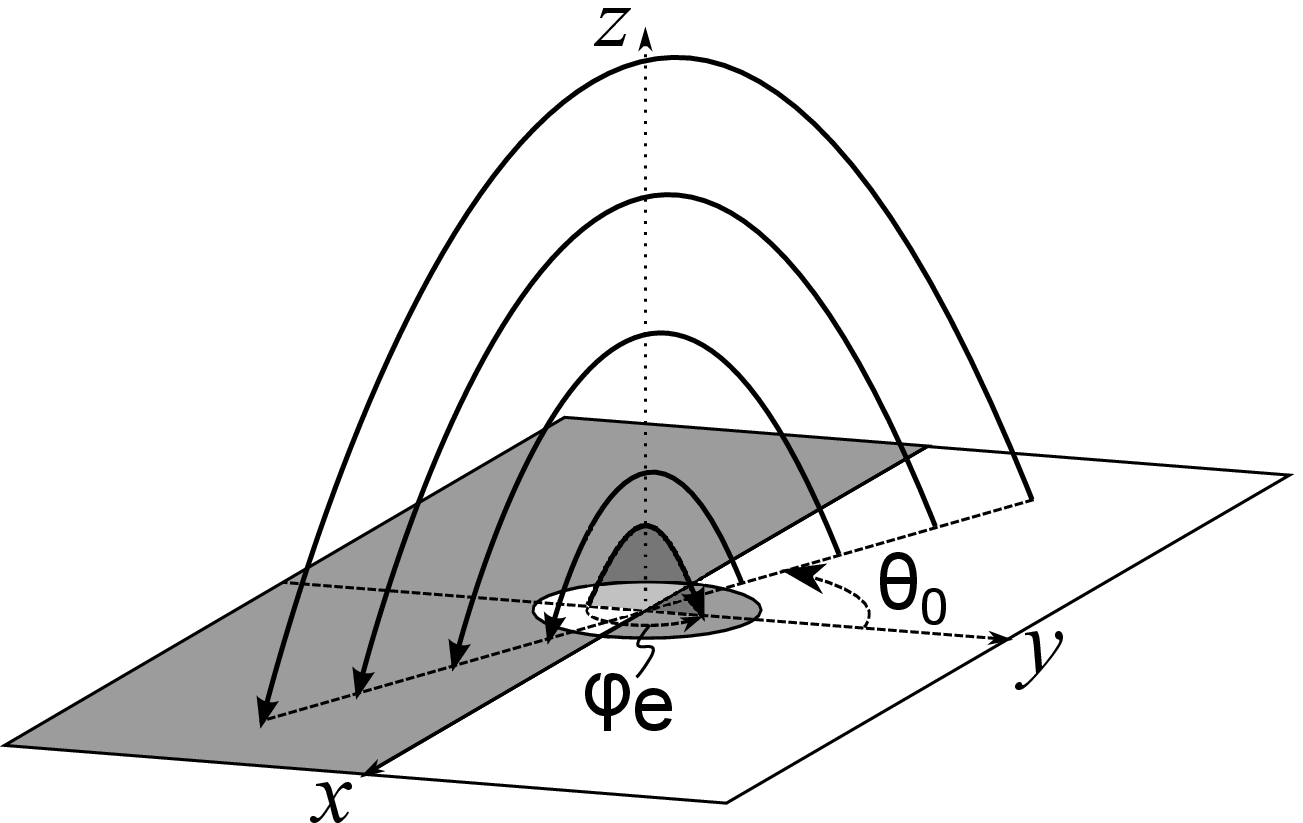} 
\caption{Illustration of simulation setup.
Curved solid lines with arrows correspond to magnetic field lines in the
preexisting force-free field and the small scale injected field,
which rotate $\theta_0$ and $\varphi_e,$
respectively, with respect to the large scale potential field.
White and gray areas on the bottom surface indicate
positive and negative magnetic polarity regions.}
\end{center}
\end{figure}
\begin{figure}[h] 
\begin{center}
\includegraphics[scale=0.9]{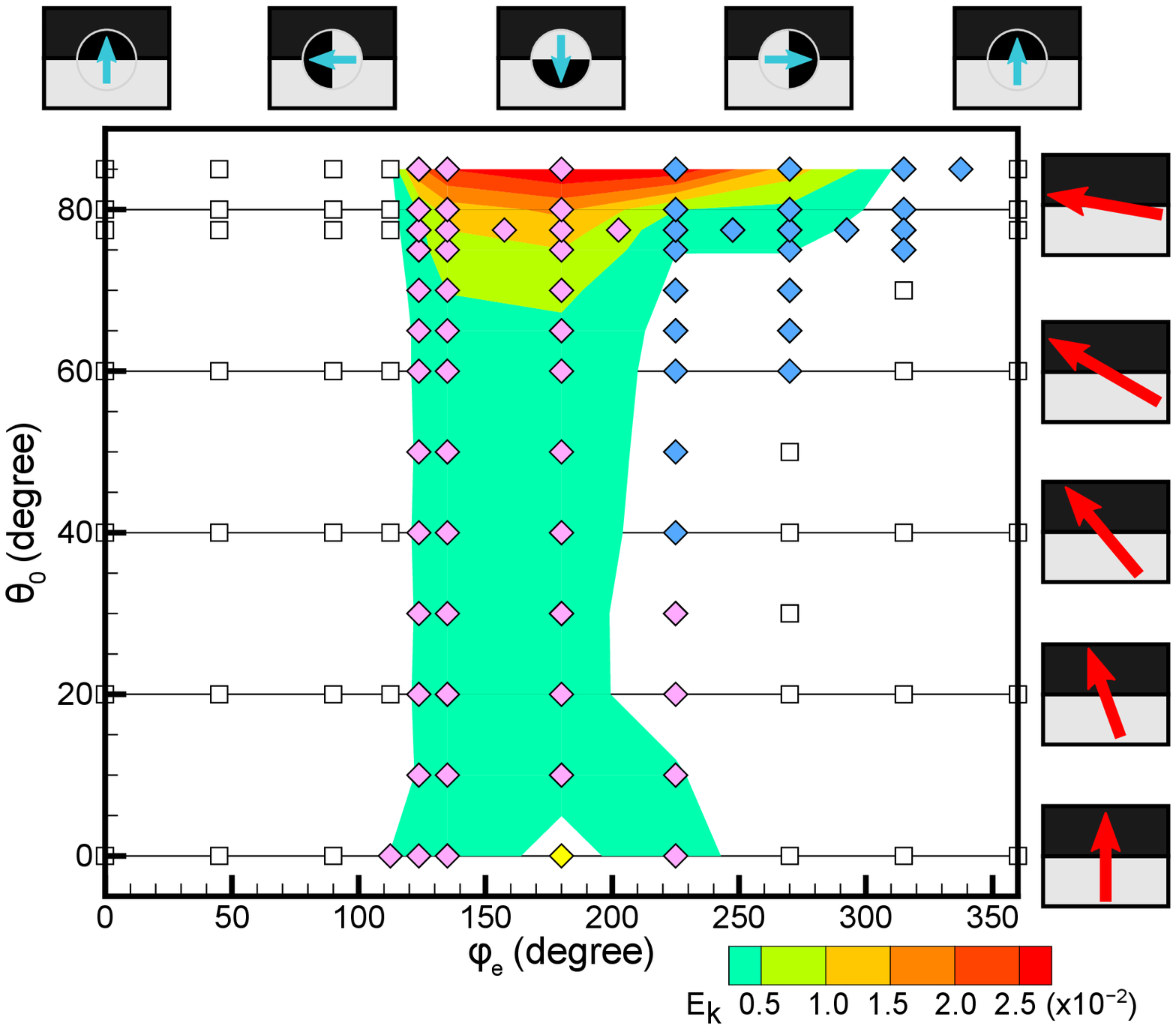} 
\end{center}
\end{figure}
\begin{figure}
\caption{Summary of simulations for $B_e = 2,$ 
$r_e = 0.13,$ $\tau_e= 20,$ $v_e = 6.7 \times 10^{-3},$ 
$y_e=0$ on parameter space of $\varphi_e$ and $\theta_0.$ 
Different marks (squares and diamonds) represent the types of dynamics, and 
contours show the maximum total kinetic energy produced by eruption ($E_k$). 
Squares indicate that 
no eruption has occurred at the corresponding parameter; 
diamonds indicate the appearance of eruptions 
at each parameter. Pink and blue diamonds indicate 
eruption-induced reconnection and 
reconnection-induced eruption processes, respectively. 
The yellow diamond corresponds to a special case in 
which the potential field collapses because of reconnection 
with the small scale injected field, 
which exhibits a completely antiparallel polarity compared with the initial 
potential field. Right hand and top 
subsets illustrate the initial sheared field and orientation 
of injected small bipole field, respectively, in 
which white and black areas indicate positive and negative polarity 
and arrows represent the 
horizontal component of the magnetic field.}
\end{figure}
\begin{figure}[h] 
\begin{center}
\includegraphics[scale=0.9]{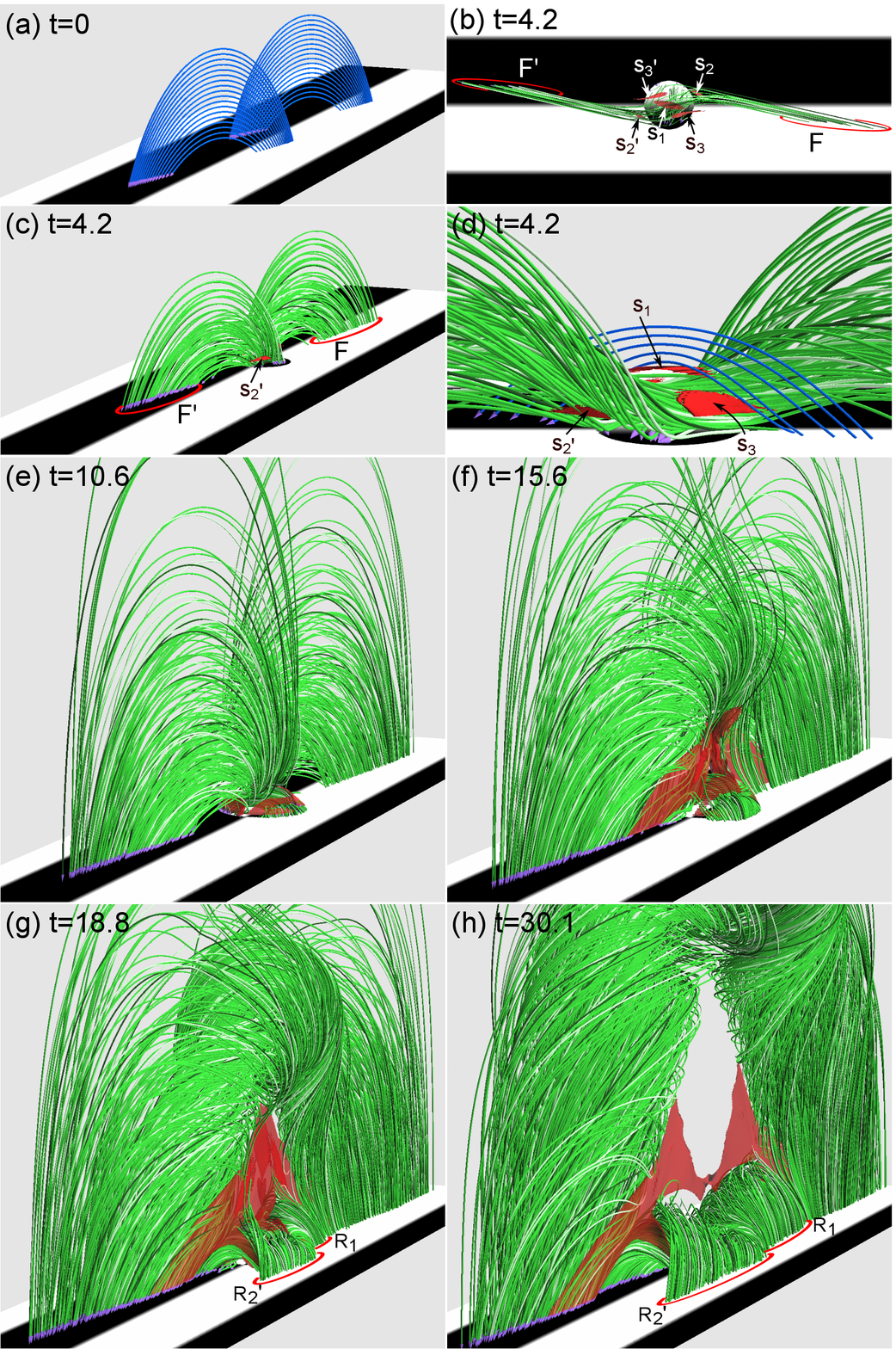}
\end{center}
\end{figure}
\begin{figure}[h] 
\caption{Simulation result for $\varphi_e = 180^{\circ}$ and
$\theta_0 = 77.5^{\circ}$, in which the OP-type
of magnetic structure causes the eruption-induced reconnection dynamics. 
Each subset represent a birds eye view (a, c, e-h), top view
(b), and enlarged side view (d) of the magnetic field at different times.
Green tubes represent magnetic field lines with connectivity that differs from 
the initial state. 
Selected magnetic fields in the initial state and 
those retaining the initial connectivity
are plotted by blue tubes in (a) and (d), respectively.
Red contours correspond to intensive current layers in which 
$ | \nabla \times \mathbf{B} |  > 40.$
Gray scales (white, positive; black, negative) on the bottom plane 
indicate the distribution of the $z$ component of magnetic 
field $B_z.$
Other parameters are $B_e = 2$, 
$r_e = 0.13,$ $\tau_e = 20,$ $v_e = 6.7 \times 10^{-3},$ and 
$y_e=0.$}
\end{figure}
\begin{figure}[h] 
\begin{center}
\includegraphics[scale=0.9]{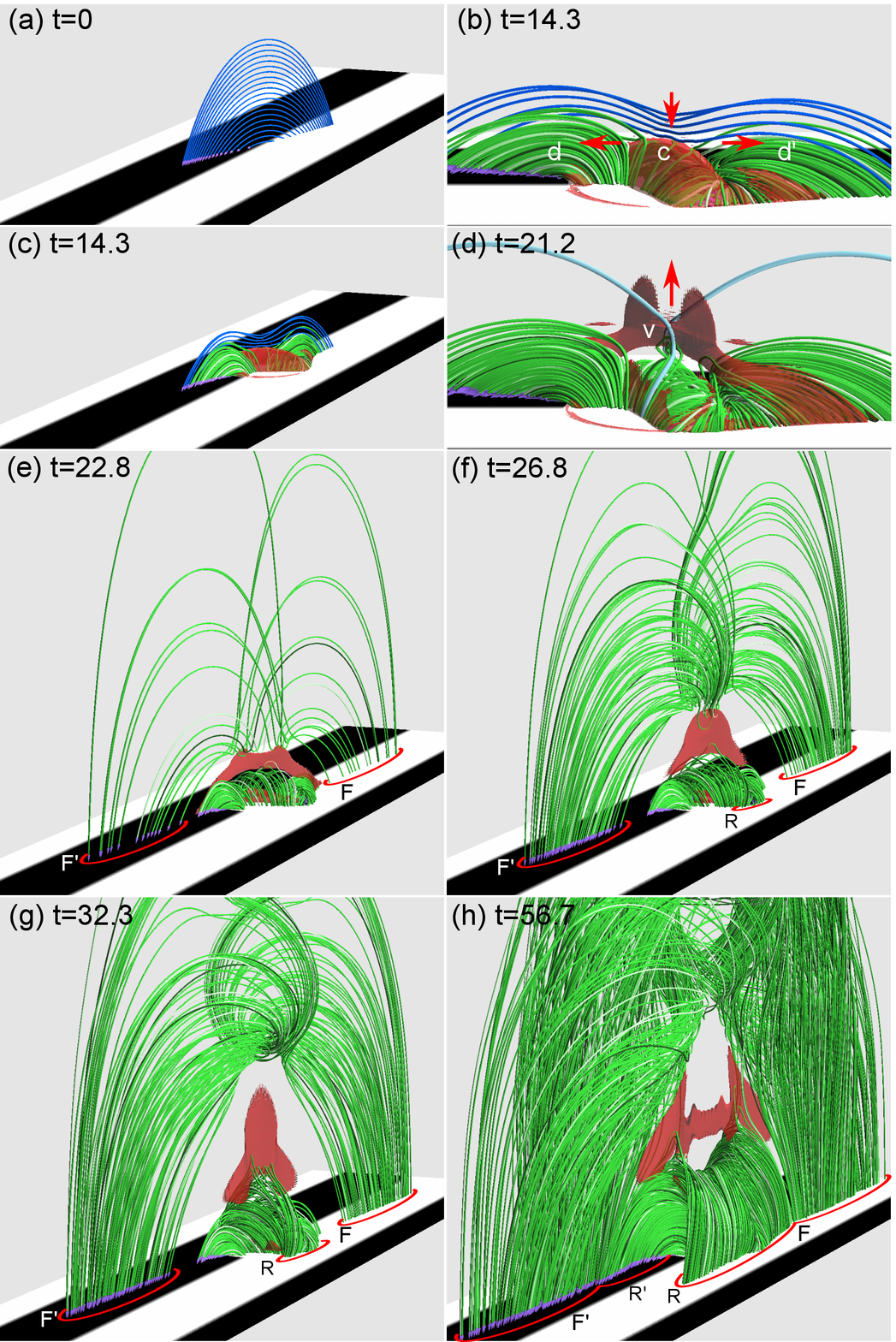} 
\end{center}
\end{figure}
\begin{figure}[h]
\caption{Simulation result for $\varphi_e = 270^{\circ}$ and
$\theta_0 = 77.5^{\circ}$, in which the RS-type of
magnetic structure 
causes the reconnection-induced eruption dynamics. 
Each subset represents a birds eye view (a, c, and e-h)
and enlarged side views (b and d) of the magnetic field at different times.
Green tubes represent magnetic field lines with 
connectivity that differs from the initial state. 
Selected magnetic fields in the initial state 
and those retaining the initial connectivity
are plotted by blue tubes in (a) and blue and sky-blue tubes in
(b-d), respectively. Red arrows in (b-d) represent the typical flow
directions. The format and other parameters are
same as those described in Fig.\ 3.}
\end{figure}
\begin{figure}[h] 
\begin{center}
\includegraphics[scale=0.85]{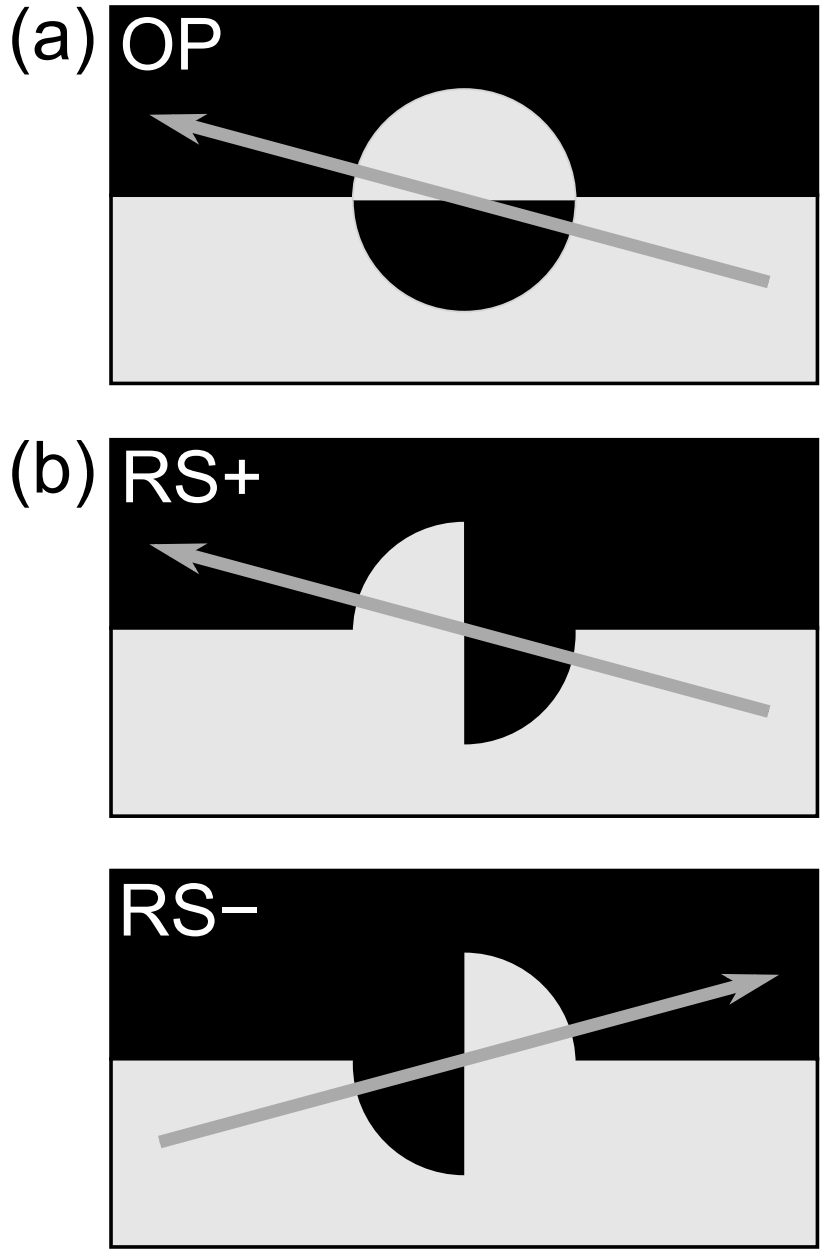}
\caption{Schematic diagram of top view of (a) OP and 
(b) RS-type 
magnetic structures. Arrows indicate the large scale 
sheared magnetic field, and black and white parts
correspond to the negative and positive polarity regions.
The chirality of PIL disturbance in the RS-type structure
is changed depending on the sign of magnetic shear;
RS+ for positive shear (counterclockwise rotation) and 
RS$-$ for negative shear (clockwise rotation). }
\end{center}
\end{figure}
\begin{figure}[h] 
\begin{center}
\includegraphics[scale=0.85]{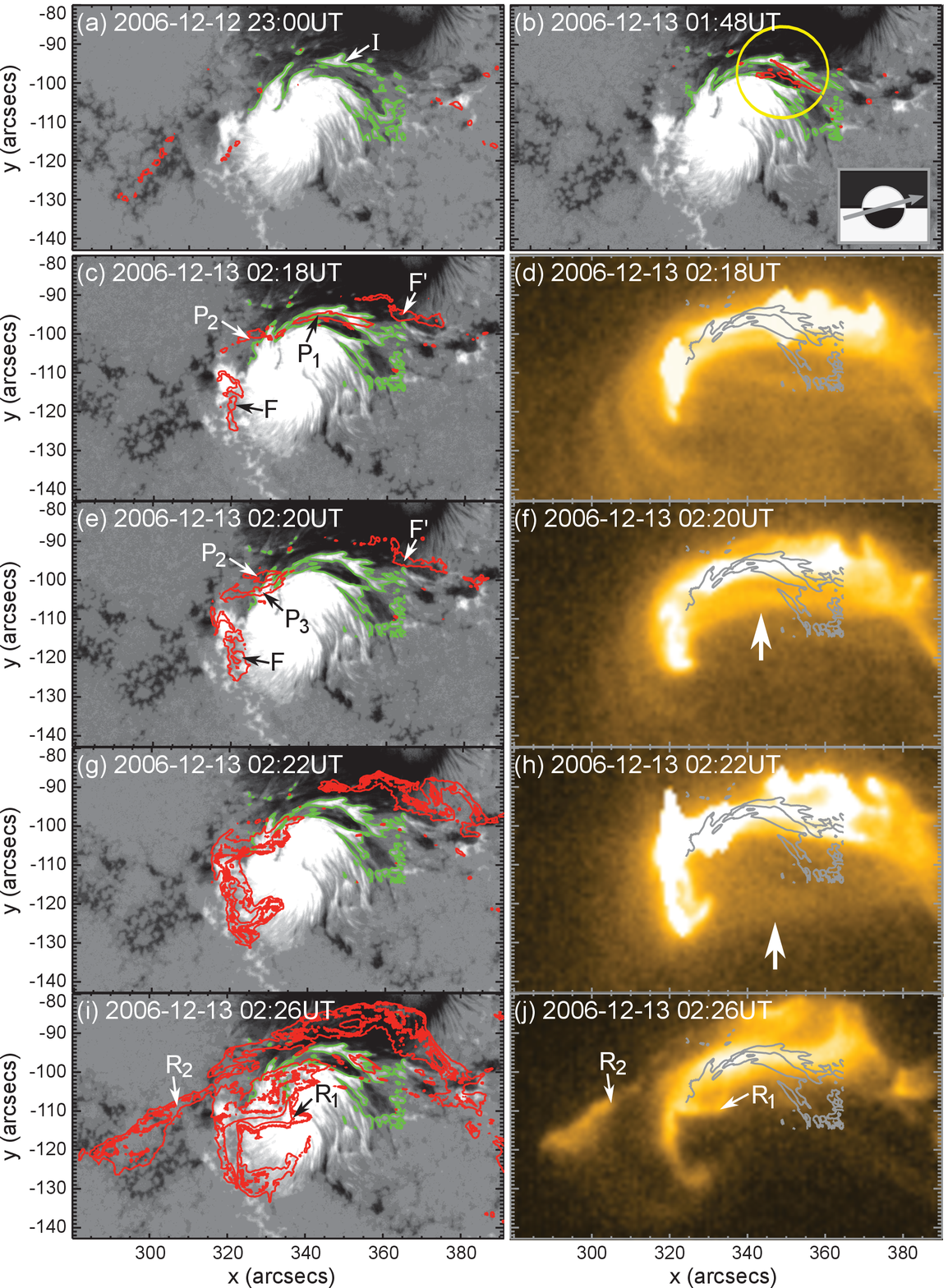}
\end{center}
\end{figure}
\begin{figure}
\caption{Onset processes for solar flares in  
AR NOAA 10930 observed by Hinode/SOT 
(a-c, e, g, and i), and soft X-ray images obtained with 
the Hinode/XRT (d, f, h, and j).
Gray scale in the SOT data 
represents the Stokes-V/I signal
of the Fe I 6302 \AA\ line, 
which indicates circular polarization corresponding 
to the line-of-sight component of the magnetic field
(black, negative; white, positive).
Red contours indicate the emission of the Ca II H line. 
Green lines on the SOT data
and gray lines on the XRT data represent the polarity 
inversion line (PIL)
in the central region. 
Observation times are shown for each subset.
Subset at the lower right corner of (b) illustrates
the typical OP-type magnetic structure with clockwise 
rotating sheared field.}
\end{figure}
\begin{figure}[h] 
\begin{center}
\includegraphics[scale=0.9]{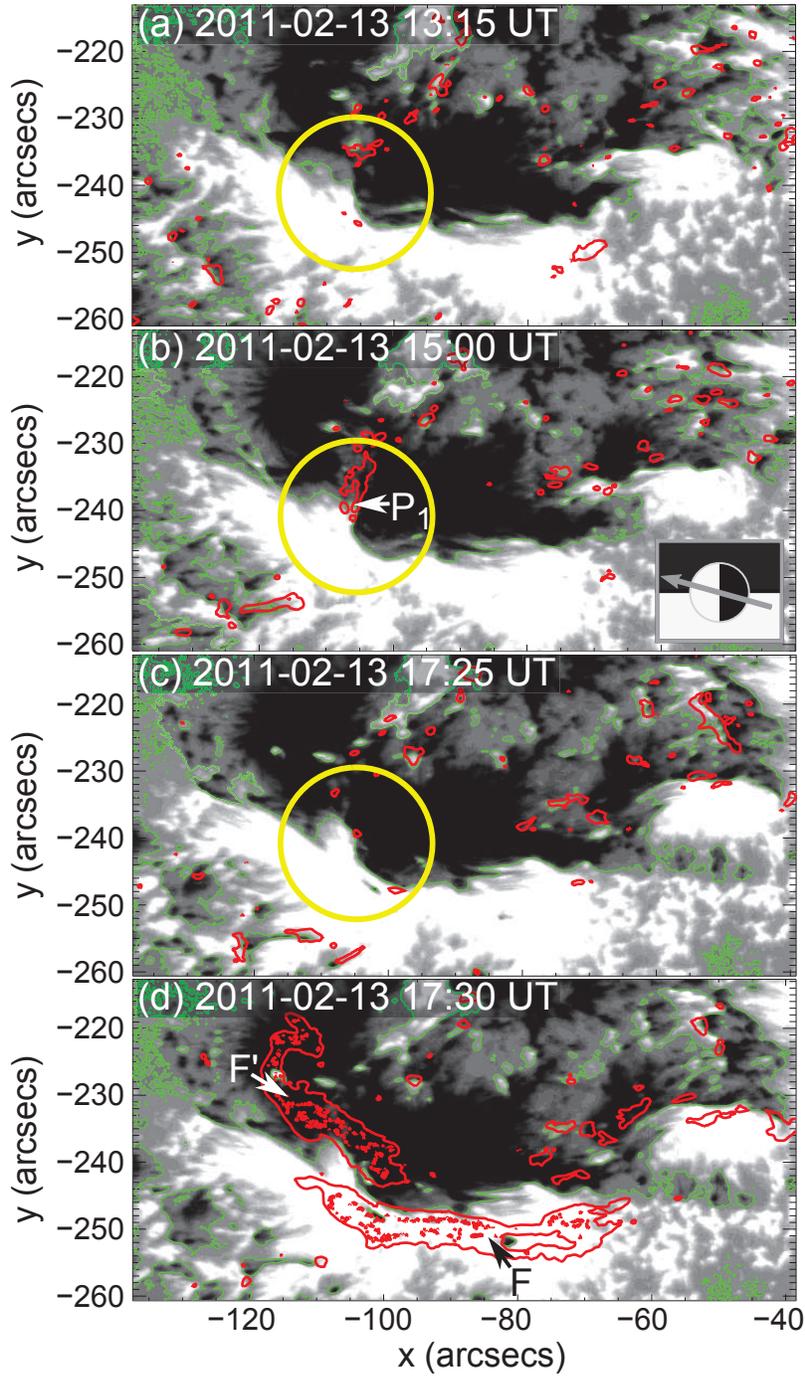}
\end{center}
\caption{Onset processes for solar flares in 
AR NOAA 11158 observed by Hinode/SOT.
Gray scale (black, negative; white,
positive) represents the Stokes-V/I signal 
in the Na I 5896 \AA\ line.
Red contours indicate the emission of the Ca II H line, and 
green lines represent the polarity inversion line (PIL).  
Observation times are shown for each subset.
Subset at the lower right corner of (b) illustrates
the typical RS structure with counterclockwise rotating sheared
field.}
\end{figure}
\begin{figure}[h] 
\begin{center}
\includegraphics[scale=1.0]{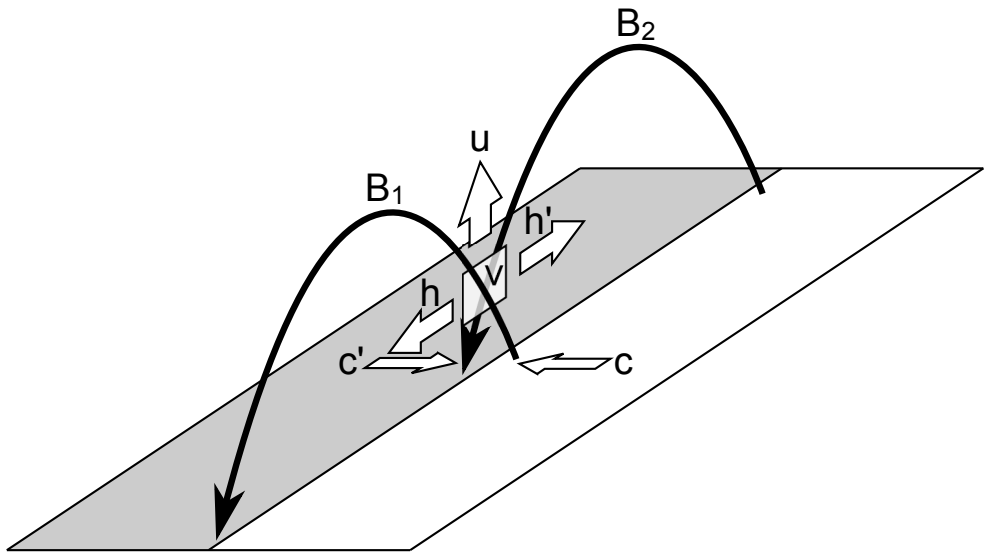}
\end{center}
\caption{Illustration of the three different
flows {\sf u}, {\sf h}-{\sf h'}, and {\sf c}-{\sf c'}, 
which may cause the formation of a vertical
current sheet {\sf v} and magnetic reconnection
between two sheared magnetic loops {\sf B$_1$} and
{\sf B$_2$} on the polarity inversion line (PIL).}
\end{figure}
\end{document}